\documentclass[twocolumn,showpacs,preprintnumbers,amsmath,amssymb,floatfix,prb]{revtex4}

\usepackage{graphicx}
\usepackage{dcolumn}
\usepackage{bm}

\begin{document}

\preprint{APS/123-QED}

\title{Variational local moment approach: from  Kondo effect \\to Mott transition in correlated electron systems} 

\author{Anna Kauch}
\email{Anna.Kauch@physik.uni-augsburg.de}
 \affiliation{Theoretical Physics III, Center for Electronic Correlations and
Magnetism, Institute of Physics, University of Augsburg, D-86135
Augsburg, Germany} 
\author{Krzysztof Byczuk}%
 
\affiliation{%
Institute of Theoretical Physics,
 University of Warsaw, 
ul. Ho\.za 69,
 PL-00-681 Warszawa,
 Poland
}%


\date{\today}

\begin{abstract}
{\sf The variational local moment approach (VLMA) solution of the single impurity Anderson model is presented. It generalizes the local moment approach of Logan {\it et al}. by invoking the variational principle to determine the lengths of local moments and orbital occupancies. We show that VLMA is a comprehensive, conserving and thermodynamically consistent approximation and treats both Fermi and non-Fermi liquid regimes as well as local moment phases on equal footing. We tested VLMA on selected problems. We solved the single- and multi-orbital impurity Anderson model in various regions of parameters,  where different types of Kondo effects occur. The application of VLMA as an impurity solver of the dynamical mean-field theory, used to solve the multi-orbital Hubbard model, is also addressed.}
\end{abstract}

\maketitle

\section{\label{sec:level1}Introduction}

The minimum at low temperatures in the resistance of metals with magnetic impurities,\cite{kondo_first} the so-called Kondo effect, is related to generic singularities in the many-body perturbation expansion.\cite{kondo} Many efforts have been put to developing a complete theory of the Kondo effect, which would be free from such infinities.\cite{anderson,schrieffer_wolff,wilson,krishnamurty,ts_wiegman,andrei_rmp} Below a typical crossover temperature (Kondo temperature) there is a many-body screening of the localized magnetic moment by extended electrons with spin.\cite{hewson,coleman} As a result, a many-body singlet state is formed. The local moment is shielded or confined and  the properties of such a many-body system are well represented by the local Fermi liquid picture.\cite{nozieres} This is a strong-coupling regime. Above the Kondo temperature the localized magnetic moment is decoupled from the electrons giving a Curie contribution to the magnetic susceptibility. This 
 is a local moment regime. Now, the Kondo effect attracts attention elsewhere:\cite{kondo_viewpoint} in classical\cite{kosterlitz_thouless,gogolin_tsvelik} and quantum phase transitions,\cite{sachdev_qpt} in quantum dots and nanostructures,\cite{kondo_revival} in strongly correlated electrons,\cite{dmft_phys_today} and also in  quark confinement and asymptotic freedom in  hadronic matter.\cite{wilson_nobel,varma_confinement,coleman_confinement}

Traditionally, the Kondo effect is studied either within the Kondo model,\cite{kondo} where the extended electrons and the localized magnetic moment interact via the exchange antiferromagnetic coupling, or within the single impurity Anderson model (SIAM),\cite{anderson} where the extended electrons hybridize with the localized electrons and the latter interact via on-site Coulomb-like interaction. 

The Kondo model and the SIAM with linear dispersion relations have been solved exactly by the Bethe ansatz technique.\cite{ts_wiegman,andrei_rmp} This powerful method is not, however, applicable in general cases. Other analytical, asymptotically exact methods, i.e. conformal field theory and bosonization, are also limited to specific dispersion relations. In many practical cases, e.g. in modeling transport in nanostructures\cite{kondo_revival} or in modeling correlated electron systems within the dynamical mean-field theory (DMFT),\cite{dmft_phys_today} the Kondo model or the SIAM would have to be solved for arbitrary dispersion relations, exchange interaction, and/or hybridization functions. This requirement is achieved within numerically exact methods, e.g. in quantum Monte Carlo (QMC)\cite{d_qmc,ct_qmc} or in numerical renormalization group (NRG).\cite{krishnamurty,nrg_review} The computational cost in time and memory increases, however, very fast at lowest temperatures
  or in systems with many channels and orbitals. Therefore, all those exact methods are of limited use in modeling real systems.

Different analytical, approximate methods have also been developed. In fact, our understanding of the Kondo effect arrived from renormalization group approaches,\cite{anderson_poor,kehrein_flow,functional_rg}  where high energy modes are successively removed and low energy couplings are renormalized  accordingly. Another class of approximate methods are those based on different perturbation expansions.\cite{zlatic1,Yosida,nca} They do not reproduce either the exponential Kondo energy scale or the local Fermi liquid properties at energies below the Kondo energy scale. These drawbacks can  be partially  cured within sophisticated resummations of Feynman diagrams.\cite{janis1,rubtsov}

In recent years a local moment approach (LMA) has been put forward by D.~Logan and collaborators.\cite{logan1,logan2,logan3,logan4} Within this analytical, approximate  method applied to the SIAM one recovers: i) the exponential Kondo energy scale, ii) low-energy local Fermi liquid properties, and iii) high energy charge fluctuations, correctly. The LMA starts with an unrestricted Hartree-Fock (UHF) perturbation expansion, derived for each direction of the local moment, and afterwards the symmetry restoration is performed.\cite{verwoerd,logan1} Physically relevant observables are derived from averaging perturbation results over different local moment orientations. Finally, a free parameter, i.e. the length of the local moment, is determined via imposing the local Fermi liquid conditions on the ground state. Since the LMA has been proved to reproduce several exactly known results and to recover correctly the properties of the SIAM at all energy scales\cite{logan1,logan2,logan3,logan4} it seems to be an excellent method of choice in studying properties of nanosystems\cite{kondo_revival} and systems with correlated electrons.\cite{dmft_review} Indeed the LMA has been used in studying Kondo insulators and heavy-fermions,\cite{logan_pam1,logan_pam2,logan_pam3,logan_pam4} impurity systems with many orbitals,\cite{logan_multi} and SIAM in the presence of magnetic field.\cite{slma}

In this paper we generalize the LMA of D.~Logan {\it et al}.\cite{logan1,logan2,logan3,logan4} by using the variational principle in determining the length of the local magnetic moment. Such approach is motivated by the very general principle that the best approximate solution must minimize the ground state (free) energy. It is free from any bias toward the local Fermi liquid properties. Therefore, the variational local moment approach (VLMA) is applicable to any system with a magnetic impurity, not necessarily that being in the local Fermi liquid fixed point. In particular, VLMA can be used in studying Kondo effects in multi-orbital and/or multi-channel SIAM, where local Fermi or non-Fermi liquid ground states are expected.\cite{cox_zawadowski} It can be extended to finite temperatures, where local Fermi liquid properties are only approximately satisfied. VLMA can also be applied to solving the impurity problems inside the DMFT and its cluster extensions.\cite{dual_fermion,dca,cdmft} This allows for unbiased studying of metal-insulator transitions in correlated electron systems in various configurations. In this paper we present a few applications of the VLMA. 

In the following we define the multi-orbital SIAM in Sec.~II and next formulate the VLMA method in Sec.~III. Here we also provide arguments that VLMA is a conserving and thermodynamically consistent approximation in the Kadanoff-Baym sense\cite{Kadanoff}. To show the usefulness of this method, the VLMA is applied to one- and two-orbital SIAM in Sec.~IV and to correlated electron problems within the DMFT in Sec.~V. Final conclusions and outlooks are in Sec.~VI.


\section{Multi-orbital Anderson model}

The VLMA is developed explicitly for solving a multi-orbital single impurity Anderson model, which is defined first. The Hamiltonian of the multi-orbital SIAM is given by
\begin{eqnarray}
H=\sum_{\bf{k}\sigma \alpha} \epsilon_{\bf k} c_{{\bf k}\alpha\sigma }^{\dagger}
c_{{\bf k}\alpha\sigma }^{\phantom\dagger} + \sum_{{\bf k}\sigma\alpha \beta} \!\!\left( V_{{\bf k}\alpha \beta}
d_{\alpha\sigma}^{\dagger} c_{{\bf k}\beta\sigma}^{\phantom\dagger} + h.c.
\right) +  \nonumber  \\
  \sum_{\alpha\sigma}\left(\epsilon_{\alpha}+\frac{1}{2}U_{\alpha}
n_{\alpha-\sigma}\right) n_{\alpha\sigma}+ \nonumber  \\
\frac{1}{2} \sum_{\sigma\sigma'}\sum_{\alpha\neq\beta}\left( U'_{\alpha\beta}
-J_{\alpha \beta}\delta_{\sigma\sigma'}\right)
n_{\alpha\sigma}n_{\beta\sigma'}, \quad
\label{anderson_multi}
\end{eqnarray}
where $d_{\alpha\sigma}^{\dagger}$ creates an impurity electron with spin $\sigma=\uparrow, \downarrow$ and  orbital quantum number $\alpha=1,2,\ldots,N$, $n_{\alpha\sigma}= d_{\alpha\sigma}^{\dagger}d_{\alpha\sigma}$ is the number operator of impurity electrons, and $c_{{\bf k} \alpha \sigma}^{\dagger}$  creates a conduction electron with spin $\sigma$ in a band~$\alpha$. We assume the same number of orbitals (bands) for local and conduction electrons. The Hamiltonian \eqref{anderson_multi} contains local interactions $U_{\alpha}$ between impurity electrons on the same  orbital and local interactions $U'_{\alpha \beta}$ between impurity electrons on different orbitals. The Ising part of the exchange interaction $J_{\alpha \beta}$ is also included. The single particle energies of conduction electrons $\epsilon_{\bf k }$ and the hybridization between impurity and conduction electrons, represented by matrix elements $V_{{\bf k}\alpha \beta}$ in \eqref{anderson_multi}, can be 
 combined into a hybridization function  $\mathbf{\Delta}(i\omega_n)$. It is a matrix in orbital space with matrix elements given by
\begin{equation}
[\mathbf{\Delta}]_{\alpha\beta}(i\omega_n) \equiv \Delta_{\alpha \beta}(i\omega_n) = \sum_{{\bf k} \gamma}\frac{V_{\bf k}^{\alpha \gamma}V_{\bf k}^{*\gamma \beta}}{i\omega_n + \tilde{\mu}-\epsilon_{\bf k}},
\label{hybridization}
\end{equation}
where $\omega_n$ are the fermionic Matsubara frequencies with $\beta=1/(k_B T)$, and $k_BT$ is temperature in energy units, whereas $\tilde{\mu}$ is the chemical potential.

\section{Variational local moment approach}

Starting from the LMA of Logan {\it et al}.,\cite{logan1,logan2,logan3,logan4} we formulate its variational version, i.e. the VLMA, in order to approximately solve the Hamiltonian \eqref{anderson_multi} with arbitrary parameters. The VLMA consists of four steps:
\begin{itemize}
\item[i)] finding the unrestricted Hartree-Fock (UHF) solution of \eqref{anderson_multi},
\item[ii)] performing a renormalized perturbation expansion around each of the UHF solutions,
\item[iii)] restoring the symmetry of the solution,
\item[iv)] variationally determining the lengths of local moments and orbital occupancies.
\end{itemize}
Next we discuss each step in detail.

\subsection{Unrestricted Hartree-Fock approximation}

The VLMA begins with the UHF solution of the SIAM in Eq.~\eqref{anderson_multi}. To set the notation and explain basic ideas we first consider  the one-orbital case, i.e. $\alpha=\beta=1$, $U_{\alpha}=U$, $\epsilon_{\alpha}=\epsilon_1$. Then the UHF local Green's function $\tilde{G}_{\sigma}(i\omega_n)$ (in the Matsubara representation) is given by\cite{anderson}
$$
\tilde{G}_{\sigma}(i\omega_n) = \frac{1}{i\omega_n +\tilde{\mu}- \epsilon_1- \Delta(i\omega_n) - U \langle n_{-{\sigma}} \rangle}.
$$
where $\Delta(i\omega_n)=\Delta_{11}(i\omega_n)$ in Eq.~\eqref{hybridization} and $\langle n_{\sigma} \rangle$ is the average occupation number per spin. By introducing the total average occupancy of the impurity $
n\equiv\langle n_{\uparrow} \rangle +\langle n_{\downarrow} \rangle
$ and the local magnetic moment
\begin{equation}
\mu \equiv \langle n_{\uparrow} \rangle -\langle n_{\downarrow} \rangle,
\end{equation}
we write the UHF solution in the form
$$
\tilde{G}_{\sigma}(i\omega_n) = \frac{1}{i\omega_n +\tilde{\mu}- \epsilon_1- \Delta(i\omega_n) - \frac{1}{2} U n - \sigma \frac{1}{2} U \mu},
$$
with $\sigma = \pm 1$ for spin up/down. The UHF solution depends on the value of the local moment $\mu=\mu_{HF}$, which has to be determined self-consistently. It turns out that there exists a critical value $U_c$ such that for $U>U_c$ the UHF local moment $\mu_{HF}\neq 0$.\cite{anderson} The sign (direction along the quantization axis) of the local moment is either positive or negative. The choice of one from the two possibilities breaks the symmetry of the solution with respect to the symmetry of the SIAM. This distinguishes two broken symmetry  solutions, which we label by $A$ for $\mu>0$ and by $B$ for $\mu<0$, i.e.
\begin{equation}
\tilde{G}^{A/B}_{\sigma}(i\omega_n) = \frac{1}{i\omega_n +\tilde{\mu}- \epsilon_1- \Delta(i\omega_n) - \frac{1}{2} U n \mp \sigma \frac{1}{2} U |\mu|}.
\label{uhf_green}
\end{equation}
The Dyson equation\cite{fetter}
\begin{equation}
\tilde{G}^{A/B}_{ \sigma}(i\omega_n) =\frac{1}{G^0 (i\omega_n)^{-1}-\tilde{\Sigma}^{A/B}_{\sigma} }, \nonumber 
\end{equation}
where $G^0 (i\omega_n)^{-1} =i\omega_n +\tilde{\mu}- \epsilon_1- \Delta(i\omega_n) $ is the inverse of the non-interacting Green's function, defines two UHF self-energies  $\tilde{\Sigma}^{A/B}_{\sigma}= U n/2 \pm \sigma U \mu/2$.

In the multi-orbital SIAM with $N$ orbitals the UHF solution depends on $2N$ parameters: $N$ local moments  $\mu_{\alpha}\equiv\langle n_{\alpha\, \uparrow} \rangle - \langle n_{\alpha\, \downarrow} \rangle$ and $N$ occupancies of the orbitals  $n_{\alpha}\equiv\langle n_{\alpha\, \uparrow} \rangle + \langle n_{\alpha\, \downarrow} \rangle$. In the UHF  approximation the local moments have non-zero values above critical values of the  interactions.\cite{coqblin} Similarly to the one-orbital case, the UHF solution with non-zero local moments breaks the symmetry of the Hamiltonian. In the  case of non-degenerate orbitals and zero magnetic field the UHF solution is doubly degenerate, i.e.  solutions with  $\{\mu_{\alpha}\}\equiv\{\mu_1,\mu_2,\ldots,\mu_N \}$ and $\{-\mu_{\alpha}\}$, where all signs are opposite,  have the same energy. These two equivalent solutions we denote again with  $A$ and $B$, respectively. Hence, there are two Green's functions $\tilde{\mathbf{G}}^{A/B}_{\sigma}(i\omega_n)$, which are now 
 matrices in the orbital space. The matrix Dyson equation
$$
 \tilde{\mathbf{G}}^{A/B}_{\sigma}(i\omega_n)^{-1}= {\mathbf{G}}^0(i\omega_n)^{-1} - \tilde{\mathbf{\Sigma}}^{A/B}_{\sigma}
$$
defines the two matrix self-energies $\tilde{\mathbf{\Sigma}}^{A/B}_{\sigma}$. The matrix elements of the  self-energies are given by\cite{coqblin}
\begin{eqnarray}
\tilde{\Sigma}^{A/B}_{\alpha \beta \,\sigma }\! =\! \frac{1}{2}\! \delta_{\alpha \beta}  [ U_{\alpha} n_{\alpha}\! + \nonumber \\
 \! \sum_{\gamma\neq \alpha} \!(2 U'_{\alpha \gamma}\! -\!J) n_{\gamma} \! \mp \!  \sigma ((U_\alpha \!-\! J)\mu_{\alpha} \!+ \!J \mu) ]\equiv \delta_{\alpha \beta} \tilde{\Sigma}^{A/B}_{\alpha \,\sigma },
\label{selfenergia_hf1}
\end{eqnarray}
where $\mu = \sum_{\alpha} \mu_{\alpha}$ is the length of the total impurity magnetic moment. The UHF self-energy is diagonal in the orbital index $\alpha$. Therefore, if the hybridization function \eqref{hybridization} is diagonal, the UHF propagators are also diagonal in the orbital indices, i.e.
$$
 \tilde{{G}}^{A/B}_{\alpha \beta\,\sigma}(i\omega_n)=\delta_{\alpha \beta}\tilde{{G}}^{A/B}_{\alpha \,\sigma}(i\omega_n).
$$  
Only this diagonal case is considered here.

\subsection{Symmetry restoration}

Before discussing technical details of the perturbation expansion we first explain the idea of symmetry restoration. Since there are always two equivalent UHF solutions, the higher order in $U_{\alpha\beta}$ and $J_{\alpha\beta}$ perturbation expansion around the mean-field solution is not self-evident. We perform the renormalized perturbation expansion (RPE) around each of the  UHF solutions and then restore the symmetry. Following D. Logan,\cite{logan1,logan2,logan3,logan4} we formulate a symmetry restoring ansatz in the multi-orbital case as follows
\begin{equation}
\mathbf{G}_{\sigma}(i\omega_n; \{\mu_{\alpha}\})= \frac{1}{2} \left ( \mathbf{G}^{A}_{\sigma}(i\omega_n; \{\mu_{\alpha}\})+
  \mathbf{G}^{B}_{\sigma}(i\omega_n; \{\mu_{\alpha}\})\right ). 
\label{multi_ansatz}
\end{equation} 
Note, that the two RPE local Green's functions  $\mathbf{G}_{\sigma}^A$ and  $\mathbf{G}_{\sigma}^B$ are related by  $\mathbf{G}_{\sigma}^A(i\omega_n;\{\mu_{\alpha}\})=\mathbf{G}_{\sigma}^B(i\omega_n;\{-\mu_{\alpha}\})$. The result (\ref{multi_ansatz}) depends on the values of all local moments $\mu_{\alpha}$ and, in the general case, also on all local orbital occupancies $n_{\alpha}$. These are parameters of the method, which are  not equal to their corresponding UHF values. These parameters are determined here via the variational principle (see Sec. IIID). 

\subsection{Renormalized perturbation expansion}

\begin{figure}
\includegraphics[width= 0.49\textwidth] {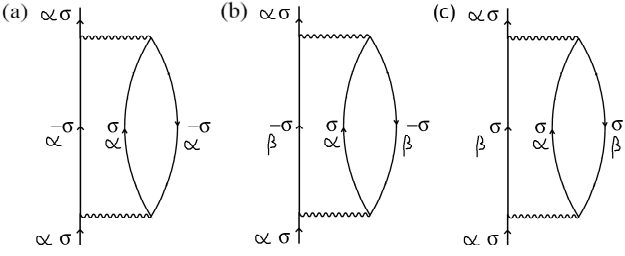}
\caption{$2^{nd}$ order Feynman diagrams : (a) a spin-flip process without the change of orbital number; (b) spin- and orbital-flip; (c) only orbital-flip. The wavy lines correspond to the local interactions: (a) $U_{\alpha}$, (b) $U'_{\alpha \beta}$, (c) $U'_{\alpha \beta}-J_{\alpha \beta}$. These diagrams contribute to the $2^{nd}$ order self-energy $\Sigma^{A(2)}_{\alpha \, \sigma}(i\omega_n)$.}
\label{multi_self}
\end{figure}

Now, we need to find  $\mathbf{G}^{A/B}_{\sigma}(i\omega_n)$. We note that the symmetry restoration step does not depend on a specific RPE scheme around UHF. However, it is important to select such an approximation that captures the processes which are essential to the Kondo problem. These are processes with the local spin-flips.\cite{kondo,anderson}  In the multi-orbital case spin-flip processes can occur together with the change of the orbital index. Such an orbital-flip is also possible without spin-flip, however. The second order Feynman diagrams describing these processes are depicted in Fig.~\ref{multi_self}.

The second order contribution to the self-energy is given by
{\small
\begin{equation}
\Sigma^{A(2)}_{\alpha \, \sigma}(i\omega_n\!)\!=\!\sum_{\beta \, \sigma'} \!\!(U_{\alpha \beta}^{\sigma \sigma'})^2  \frac{1}{\beta}\sum_{i\nu_m} \!\tilde{G}^{A\,\sigma'}_{\beta}(\!i\omega_n\!-\!i\nu_m) {^0\Pi}^{AA \,\sigma' \sigma}_{ \beta\alpha}(i\nu_m),
\label{selfenergia2}
\end{equation}
}where $\nu_m$ is the bosonic Matsubara frequency and ${\bf U}$ the local interaction matrix  with elements that have spin and orbital indices and are given by
\begin{equation}
U_{\alpha \beta}^{\sigma \sigma'}=\delta_{\alpha \beta}\delta_{\sigma -\sigma'} U_{\alpha} + (1-\delta_{\alpha \beta})(U'_{\alpha \beta} - \delta_{\sigma \sigma'} J).
\label{Udef}
\end{equation}
By construction, since $\mathbf{\Delta}(i\omega_n)$ is diagonal, the self-energy is also diagonal in orbital and spin indices. The polarization propagator ${^0\Pi}^{AA \,\sigma' \sigma}_{ \beta\alpha}(i\nu_m)$ that contributes to the self-energy in Eq.~\eqref{selfenergia2} is given by 
{\small
\begin{equation}
^0\Pi^{AA \,-\sigma\sigma}_{\alpha \beta}(i\nu_n)\!=\! -\! \frac{1}{\beta}\! \sum_{i\omega_m}\tilde{G}^{A\,-\sigma}_{\alpha }(i\omega_m)\tilde{G}^{A\,\sigma}_{\beta}(i\nu_n+i\omega_m),
\label{pi_zero}
\end{equation}
}in the transverse spin channel ($\sigma'=-\sigma$), and by
{ \small
\begin{equation*}
^0\Pi^{AA \,\sigma\sigma}_{\alpha \beta}(i\nu_n)\!=\! -\! \frac{1}{\beta}\! (\!1\!-\!\delta_{\alpha \beta}\!)\! \sum_{i\omega_m}\!\tilde{G}^{A\,\sigma}_{\alpha}\!(i\omega_m)\!\tilde{G}^{A\,\sigma}_{\beta}(\!i\nu_n\!+\!i\omega_m)
\end{equation*}
}in the longitudinal channel ($\sigma'=\sigma$). Feynman diagrams representing these expressions are plotted in Fig~\ref{multi_self}.

\begin{figure}
\includegraphics[width= 0.49\textwidth] {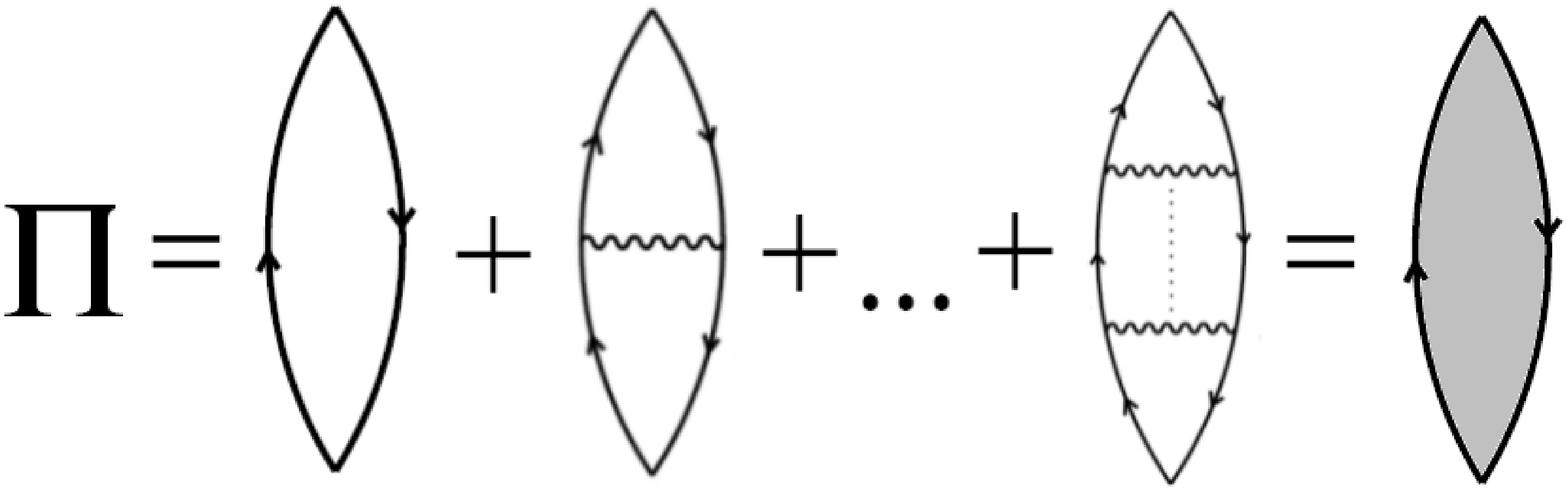}
\caption{The ladder diagrams contributing to the full polarization propagator $\mathbf{\Pi}^{AA}(i\nu_n))$. }
\label{rpa2}
\end{figure}

Finite order perturbation theory does not reproduce the Kondo exponential scale. Therefore, following Refs.~\onlinecite{logan1,logan2,logan3,logan4}, we perform the ladder summation of diagrams with spin- and orbital-flip processes to infinite order. Then we obtain the self-energy with a new polarization propagator $\mathbf{\Pi}^{AA}(i\nu_n)$. The summed series, depicted symbolically in Fig.~\ref{rpa2}, leads to a matrix expression for the polarization propagator 
\begin{equation}
\mathbf{\Pi}^{AA}(i\nu_n)= {^0\mathbf{\Pi}}^{AA}(i\nu_n)(\mathbf{1}-\mathbf{U} {^0\mathbf{\Pi}}^{AA}(i\nu_n))^{-1},
\label{pi}
\end{equation}
where $\mathbf{U}$ is the interaction matrix  \eqref{Udef}. Then the self-energy is obtained by inserting the full polarization propagator  $\mathbf{\Pi}^{AA}(i\nu_n)$ to the expression \eqref{selfenergia2}, i.e. 
{\small
\begin{equation}
\Sigma^{A}_{\alpha \, \sigma}(i\omega_n)=\sum_{\beta \, \sigma'} (U_{\alpha \beta}^{\sigma \sigma'})^2 \, \frac{1}{\beta}\sum_{i\nu_m} \tilde{G}^{A\,\sigma'}_{\beta}(i\omega_n-i\nu_m) {\Pi}^{AA \,\sigma' \sigma}_{ \beta\alpha}(i\nu_m),
\label{selfenergia3}
\end{equation}
}where we replaced  $^0\mathbf{\Pi}^{AA}(i\nu_n)$ by  $\mathbf{\Pi}^{AA}(i\nu_n)$.

The UHF Green's functions $\tilde{\mathbf{G}}^{A/B}_{\sigma}(i\omega_n)$ have the symmetry
$$\tilde{\mathbf{G}}^{A}_{\sigma}(i\omega_n)=\tilde{\mathbf{G}}^{B}_{-\sigma}(i\omega_n).$$
This leads to the relations between matrix elements of $\mathbf{\Pi}^{BB}(i\nu_n)$ and $\mathbf{\Pi}^{AA}(i\nu_n)$, i.e.
\begin{equation}
\Pi^{BB \,\sigma\sigma'}_{\alpha \beta}(i\nu_n)={\Pi}^{AA \,-\sigma -\sigma'}_{\alpha\beta }(i\nu_n).
\end{equation}
Hence, we have a useful relation between the two self-energies
\begin{equation}
\Sigma^{B}_{\alpha \, \sigma}(i\omega_n) =  \Sigma^{A}_{\alpha \, -\sigma}(i\omega_n). 
\label{aa_bb3}
\end{equation}
The frequency-dependent self-energies $\Sigma^{A/B}_{\alpha \, \sigma}(i\omega_n)$ can now be used in
 the Dyson equation  to obtain the two propagators $\mathbf{G}^{A/B}_{\sigma}(i\omega_n)$. The full solution is given by  the symmetry restoration  equation \eqref{multi_ansatz}. The solution is still dependent on the local moments $\{\mu_{\alpha} \}$ and on the orbital occupancies. They are determined next.


%

\subsection{Variational principle}

The values of the local moments are yet unknown parameters.  The physical values are found by the minimization of the ground state energy at zero temperature or the free energy at finite temperatures. This variational approach  is the main difference between the VLMA, presented here, and the LMA in Refs.~[\onlinecite{logan1,logan2,logan3,logan4}]. The general variational approach allows us to use the method for more than one orbital in the degenerate and non-degenerate cases, and also enables us to formulate the method directly at finite temperatures.\cite{kauch,kauch2,kauch_phd}

In the SIAM, the ground state energy is obtained from the one-particle local Green's function. The total ground state energy in the SIAM has two contributions: $E_{\rm bulk}$ and $E_{\rm imp}$. The energy $E_{\rm bulk}$ does not depend on the local moments $\{\mu_{\alpha} \}$. Therefore, we only need to find the minimum of $E_{\rm imp}$, which is given by\cite{ground_en}
\begin{eqnarray}
E_{\rm imp}= \frac{1} {2\pi i} \sum_{\alpha \beta \,\sigma} \oint_C d \omega \left[
  \frac {1}{2}\left(\omega + \epsilon _\alpha \right) \delta_{\alpha \beta} + \right. \nonumber \\
 \left . \frac {1}{2}\Delta_{\alpha \beta} (\omega) -\omega \frac{\partial \Delta_{\alpha \beta} (\omega)}{\partial \omega} \right]
G_{\beta \alpha}^{\sigma} (\omega),
\label{eq:ground}
\end{eqnarray}
where  $C$ is a contour in the complex frequency upper half-plane, and $G_{\beta \alpha}^{\sigma} (\omega)$  and $\Delta_{\alpha \beta} (\omega)$ are the retarded Green's  and hybridization functions, respectively. After calculating the integral, we  express the ground state energy via real functions of real frequencies, i.e.
{\small
\begin{eqnarray}
E_{\rm imp}= \frac {1}{2}\sum_{\alpha \,\sigma} \!\!\!\int_{-\infty}^{\tilde{\mu}} \!\!\!\!\frac {d \omega}{\pi} \! (\omega \!+\! \epsilon _{\alpha})\!\Im G_{\alpha \alpha}^{\sigma} (\omega)\!+ \nonumber  \\
\! \frac {1}{2}\!\sum_{\alpha \beta \,\sigma} \!\!\!\int_{-\infty}^{\tilde{\mu}} \!\!\!\!\frac {d \omega}{\pi} \!\left( \!\Im \Delta_{\alpha \beta} (\omega) \Re G_{\beta \alpha}^{\sigma} (\omega)\! + \!\Re \Delta_{\alpha \beta} (\omega) \Im G_{\beta \alpha}^{\sigma} (\omega) \!\right)\! - \nonumber \\
\! \!\sum_{\alpha \beta\,\sigma} \!\!\!\int_{-\infty}^{\tilde{\mu}}\!\!\! \frac {d \omega}{\pi}\! \left ( \! \omega \frac{\partial \Im \Delta_{ \alpha \beta} (\omega)}{\partial \omega} \Re G_{\beta \alpha}^{\sigma} (\omega) \!+ \!\omega \frac{\partial \Re\Delta_{ \alpha \beta} (\omega)}{\partial \omega} \Im G_{\beta \alpha}^{\sigma} (\omega) \!\right ).
\label{eq:energia}
\end{eqnarray}
}The physical values of $\{\mu_{\alpha} \}$ and $\{n_{\alpha} \}$ are found within the variational principle by minimizing $E_{imp}$ with respect to those parameters.

It is important to observe, that the VLMA method can be implemented by using functions on real frequency axis. This gives us a direct access to the local spectral functions $A_{\alpha\beta }^{\sigma}(\omega)=-\frac{1}{\pi} \Im G_{\alpha\beta }^{\sigma}(\omega+i0^+)$  without the necessity of performing analytical continuation. The related calculations are presented in the Appendix.

In order to use the VLMA at finite temperatures the free energy has to be minimized instead of the ground state energy. The calculation of the free energy from one-particle Green's function involves integration from zero to one over an auxiliary parameter $\lambda$ which multiplies the local interaction ${\bf U}$.\cite{fetter} Since $\lambda$ enters in a nonlinear and indirect way into the expression for the free energy, the VLMA at finite temperatures is computationally involved and has not been implemented so far. The results presented in sections IV and V have been obtained at $T=0$.

\subsection{Luttinger-Ward functional}

Any reliable approximation in the many-body theory should be thermodynamically consistent and conserving. I.e. physical observables should be the same irrespective of the way how they are determined and the microscopic conservation laws (e.g., conservation of energy, charge, spin, etc.) should be fulfilled. Within  the Kadanoff and Baym formalism,\cite{Kadanoff} we formally show that the VLMA is a thermodynamically consistent and conserving approximation. Its results are therefore reliable. In order to show this, it is enough to explicitly construct the  Luttinger-Ward functional ${\rm \Phi}[{\bf G}]$ for VLMA. Here, ${\rm \Phi}[{\bf G}]$, which  fulfills the necessary condition that $\delta {\rm \Phi}[{\bf G}]/\delta {\bf G}= {\bf \Sigma}$,\cite{Kadanoff} is given by
\begin{eqnarray*}
{\rm \Phi}[{\bf G} ] =
{\rm \Phi}[{\bf G}^A,{\bf G}^B]=\frac{1}{2}\left({\rm \tilde{\Phi}}[{\bf G}^A] + {\rm
     \tilde{\Phi}}[{\bf G}^B] \right ) +\nonumber \\
  \frac{1}{2} \mathrm{Tr} \log ( {\bf G}^A {\bf G}^B) - \mathrm{Tr} \log
\left (
   \frac{1}{2} \left ({\bf G}^A + {\bf G}^B \right ) \right),
\end{eqnarray*}
where $\mathrm{Tr}=T \sum_{i\omega_n} \mathrm{tr}$, and the symbol $\mathrm {tr}$ means the trace in the orbital and spin space,  and the functional  ${\rm \tilde{\Phi}} $ is the Luttinger-Ward functional in the ladder approximation used in Sec.~IIIC.\cite{orland_negele} It is constructed diagrammatically from ${\bf G}^{A}$ and  ${\bf G}^{B}$, respectively. The symmetry restoring constraint,  ${\bf G}=\frac{1}{2}\left( {\bf G}^A+ {\bf G}^B \right )$, is also imposed. The free energy functional $\Omega[{\bf G}]$ is now expressed via ${\rm \Phi}[{\bf G}]$ and has the following form
\begin{equation*}
\Omega[{\bf G}] = \Phi[{\bf G}] + \mathrm{Tr} \log ({-\bf G}) -
\mathrm{Tr}{\bf \Sigma G}.
\end{equation*}
The stationarity condition $\delta \Omega [{\bf G}]/\delta {\bf G}=0$ gives then the Dyson equation.

\section{Results for one-orbital SIAM}

\begin{figure}
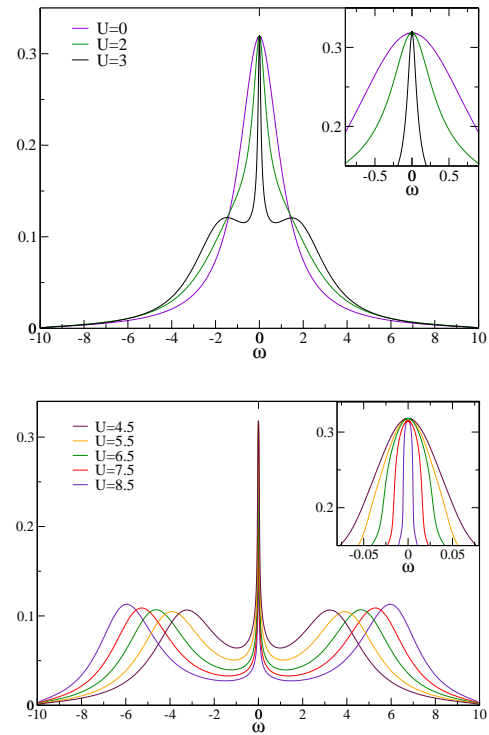

\includegraphics[width=0.35\textwidth]{fig3a.eps}
\vskip 0.5 cm
\includegraphics[width=0.35\textwidth]{fig3b.eps}
\caption{(Color online). Local spectral functions for one-orbital SIAM with semi-elliptic hybridization (half-width $D=10$, $n=1$) and different values of the interaction $U$. Inset: the energy regime around Fermi level at $\omega=0$.}  
\label{male_u}
\end{figure}

Firstly, we test VLMA in the case of one-orbital  SIAM. The spectral functions for the symmetric SIAM ($n=1$) obtained using VLMA are presented in Fig.~\ref{male_u}. Different spectral functions in this figure correspond to different values of the local interaction $U_{\alpha}=U_1=U$. The results presented here are for a semi-elliptic hybridization function $\Delta_{11}(\omega)=\Delta(\omega)$ defined by \eqref{hybridization}, whose imaginary part is given by
\begin{equation}
\Im\Delta(\omega)=\left\{
\begin{array}{ccc}
-\Delta_0 \sqrt{1-\frac{\omega^2}{D^2}} & & |\omega| < D\\
0 & & |\omega| \geq D. 
\end{array}
\right. 
\label{elipt}
\end{equation}
where $D$ is the so-called half-width and the real part $\Re\Delta(\omega)$ follows from the Kramers-Kronig relations.

\subsection{Symmetric one-orbital SIAM}

For $U=0$ the VLMA reproduces the exact result. The solution for $U\neq0$ has local Fermi liquid properties with the local spectral function pinned to its non-interacting value at the Fermi energy (cf. Fig.~\ref{male_u}) according to the Luttinger theorem.\cite{luttinger_theorem} As $U$ increases, the characteristic three-peak structure emerges with  the quasi-particle peak narrowing exponentially with $U$. The satellite peaks are the Hubbard bands.\cite{pruschke_a} Since the VLMA predicts the local Fermi liquid behavior, our results are in agreement with those obtained by the LMA if the LMA was applied to the hybridization function (\ref{elipt}).

\begin{figure}
\begin{center}
\includegraphics[width=0.3\textwidth]{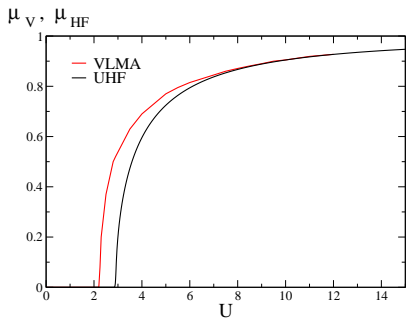}
\caption{(Color online). The value of the local moment as a function of the interaction $U$. The grey (red) curve corresponds to the VLMA solution (the minimum of $E_{imp}$ from Eq.~\eqref{eq:ground}). Black curve represents the self-consistent UHF solution $\mu_{HF}$.}  
\label{mu_od_u}
\end{center}
\end{figure}

In the symmetric SIAM with one orbital there is only one parameter, i.e. the value of the local moment $\mu$, with respect to which the ground-state energy has to be minimized. The obtained value is plotted in Fig.~\ref{mu_od_u} as a function of the interaction $U$. For small interaction, the value of the local moment is equal to zero both in the VLMA solution ($\mu_{V}$ in Fig.~\ref{mu_od_u}) and the UHF solution  ($\mu_{HF}$ in Fig.~\ref{mu_od_u}). With the increase of $U$ the values of local moments $\mu_{V}$ and $\mu_{HF}$ become different from zero at different $U$, see Fig.~\ref{mu_od_u}. For large $U$ the two solutions become exponentially close (with $\mu_{V}$ always larger than  $\mu_{HF}$). 

We note that the closeness of the variationally obtained $\mu_{V}$ to the UHF value $\mu_{HF}$ makes searching for the minimum of the ground-state energy difficult at large $U$. This is due to the divergence of $\mathbf{\Pi}(\omega)$ in Eq.~\eqref{pi} exactly when $\mu=\mu_{HF}$. Therefore, the numerical minimization of the ground state energy needs to be done carefully, especially for large $U$.

\subsection{Asymmetric one-orbital SIAM}

\begin{figure}
\includegraphics[width=0.35\textwidth]{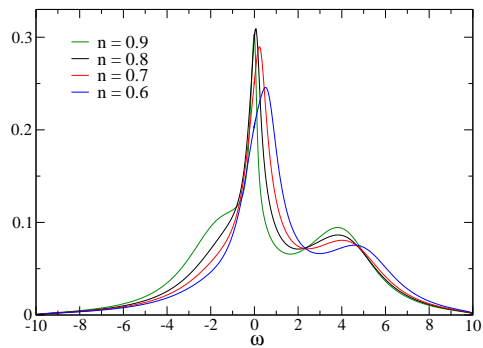}
\caption{(Color online). Local spectral functions for different impurity occupancies. All plots for $U=4$ and semi-elliptic hybridization with  $D=10$. The Fermi energy is at  $\omega=0$.}  
\label{asym1}
\end{figure}

The results for asymmetric ($n\neq1$) one-orbital SIAM are shown in Fig.~\ref{asym1}. For fillings $n<1$ the spectral weight is redistributed and, when $n$ decreases, the Kondo resonance merges with the lower Hubbard band. The values of the spectral functions at the Fermi energy are still the same as those given by the Luttinger theorem.\cite{hewson} The characteristic crossing points\cite{crossing_points} are also visible in Fig.~\ref{asym1}.

To conclude, the VLMA predicts the local Fermi liquid behavior as the solution with the lowest energy inside the manifold of different possibilities, which remains in accord to standard knowledge on the Kondo problem.\cite{hewson}

\section{Results for two-orbital SIAM}

The two-orbital SIAM is physically richer than its one-orbital counterpart. Here the interplay between direct $U_{\alpha}=U$, $U'_{\alpha \beta}=U'$ and exchange $J_{\alpha\beta}=J$  interactions and the energy difference between various impurity orbital levels are of crucial importance. In certain parameter regimes the Kondo effect is suppressed due to the quelling of fluctuations of spin and orbital degrees of freedom.\cite{cox_zawadowski} The VLMA confirms this prediction. 

\subsection{SU(4) symmetric  Kondo effect}
\begin{figure}
\includegraphics[width=0.35\textwidth]{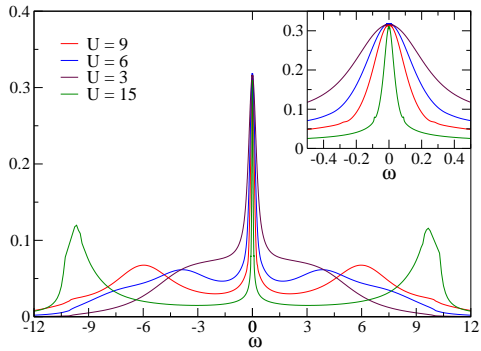}
\caption{(Color online). Local spectral functions for one of the orbitals in the half-filled ($n=2$) two-orbital SIAM for $J=0$, $U=U'$ and $\epsilon_1=\epsilon_2$. The hybridization function is semi-elliptic with half-width $D=10$, identical for both orbitals. Inset: the energy regime around the Fermi level at  $\omega=0$.}
\label{2_lma_siam}
\end{figure}

The results for the two-orbital degenerate case ($\epsilon_1=\epsilon_2$) with SU(4)\cite{cox_zawadowski} symmetry ($U=U'$, $J=0$) are presented in Fig.~\ref{2_lma_siam}. The local spectral functions have a three peak structure with the Kondo peak, which is pinned to the non-interacting value. The system remains in a local Fermi liquid fixed point up to arbitrary large values of $U$. The width of the Kondo resonance decreases exponentially with the value of $U$ as it is in the one-orbital case. However, the prefactor in the exponent, which is proportional to the number of orbitals, makes the narrowing of the Kondo peak  slower with increasing $U$ as compared to the one-orbital case. To see this we emphasize different energy scales used on horizontal axis in the insets to Figs.~\ref{male_u}~and~\ref{2_lma_siam}.

\subsection{Influence of the  exchange interaction}

In order to understand the influence of the non-zero exchange interaction (Hund's coupling) we start from the atomic limit of the two-orbital SIAM ($V_{{\bf k}\alpha \beta}=0$). In this limit at $J=0$ and $U=U'$ the impurity ground  state with two electrons is six-fold degenerate. The presence of the Hund's coupling ($J>0$) lifts this degeneracy, because the states with higher total spin ($S=1$) have lower energy. The atomic ground state is in this case two-fold degenerate ($S_z=\pm1$).

\begin{figure}
\begin{center}
\includegraphics[width=0.3\textwidth]{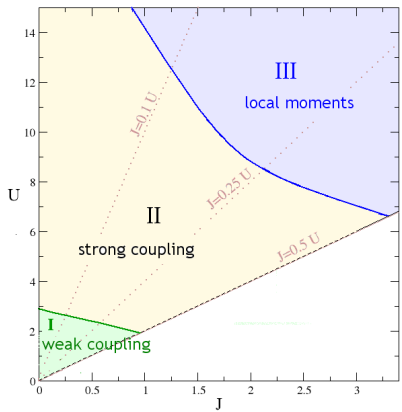}
\caption{(Color online). Phase space diagram $(U,J)$ of the ground state of the two-orbital particle-hole symmetric SIAM. The dotted (brown) lines are plotted to aid in finding points in this diagram that correspond to spectral functions plotted in  Figs.~\ref{2_lma_siam_2} and~\ref{2j_lma_siam}. The blank area corresponds to $U'<0$ (see text).}  
\label{u_od_j}
\end{center}
\end{figure}

\begin{figure}
\includegraphics[width=0.35\textwidth]{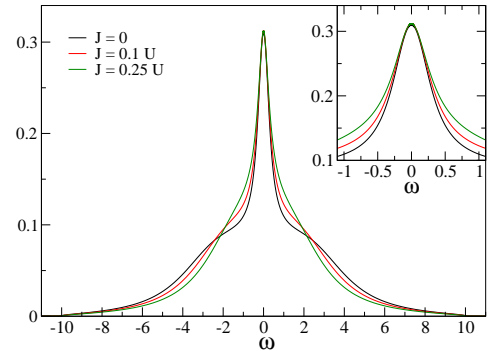}
\caption{(Color online). Local spectral functions for one of the orbitals in the two-orbital SIAM  for  $U=2$ and different values of the exchange interaction $J$ (the plots correspond to regime (I) on the diagram in Fig.~\ref{u_od_j}).  Other parameters as in Fig.~\ref{2_lma_siam}. Inset: the energy regime around the Fermi level at  $\omega=0$.}  
\label{2_lma_siam_2}
\end{figure}

\begin{figure}
\includegraphics[width=0.35\textwidth]{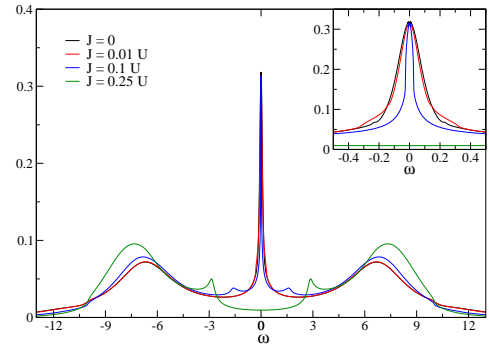}
\caption{(Color online). Local spectral functions for one of the orbitals in the two-orbital SIAM  for  $U=10$ and different values of the exchange interaction $J$ (the plots correspond to regimes (II) and (III) on the diagram in Fig.~\ref{u_od_j}).  Other parameters as in Fig.~\ref{2_lma_siam}. Inset: the area around the Fermi energy at  $\omega=0$.} 
\label{2j_lma_siam}
\end{figure}

With non-zero hybridization, the impurity spin can be flipped due to higher order processes in $V_{{\bf k}\alpha \beta}$. This leads to full or partial screening of the impurity magnetic moment. The VLMA solution of SIAM enables us to distinguish three parameter regimes with different ground states: weak coupling (I), strong coupling (II), and  local moment (III) regimes. These regimes are presented on the ground state phase diagram in Fig.~\ref{u_od_j}. We consider here the symmetric case with $U'=U-2J$. Therefore, for $J<0.5 U$, $U'$ becomes negative and this case is not analyzed here (blank area in Fig.~\ref{u_od_j}). The spectral functions for selected parameters in the  regimes (I)-(III) are presented in Figs.~\ref{2_lma_siam_2} and~\ref{2j_lma_siam}.

In the weak coupling regime (area (I) in Fig.~\ref{u_od_j}) the spins of the localized electrons are screened individually and the exchange interaction $J$ weakly influences the width of the central peak (cf. Fig.~\ref{2_lma_siam_2}). The weak coupling regime coincides with the regime where the  UHF solution gives $\mu^{HF}_1=\mu^{HF}_2=0$ (note however, that the values of the local moments in the VLMA are finite here). When $U$ increases and when the UHF solution gives non-zero values of $\mu^{HF}_1$ and $\mu^{HF}_2$, the Kondo resonance narrows and its width strongly depends on  $J$ (cf. Fig.~\ref{2j_lma_siam}). This is the strong coupling regime (II) in Fig.~\ref{u_od_j}. In the local moment regime, (area (III) in Fig.~\ref{u_od_j})  the Kondo resonance is absent. Now, the electrons, which have only spin $1/2$, cannot fully screen the spin $1$ on the impurity. The disappearing of the Kondo resonance gives rise to a critical line $J_c(U)$ that separates the strong coupling (II) and local moment (III) regimes (cf. Fig.~\ref{u_od_j}).

Finally we note that the local moment regime is absent in the two-band two-orbital SIAM with the Heisenberg exchange  interaction.\cite{hund_hubbard} Hence, the validity of the diagram presented here is restricted to the Ising-type exchange interaction.


\begin{figure}
\begin{center}
\includegraphics[width=0.3\textwidth]{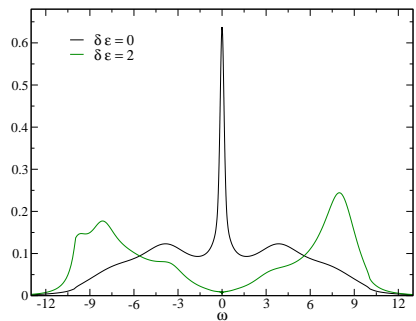}
\caption{(Color online) Local spectral functions in case of degenerate  $\delta \epsilon = 0 $ and non-degenerate $\delta \epsilon = 2 $ two orbitals in the half-filled case $n=2$ and $U=6$, $J=0$; Fermi energy is at $\omega=0$.}  
\label{2_as_as_siam}
\end{center}
\end{figure}
\subsection{Non-degenerate orbitals}

The case of non-degenerate orbitals ($\delta \epsilon=\epsilon_1-\epsilon_2\neq0$) is numerically more involved because of the increased number of parameters with respect to which the ground state energy \eqref{eq:ground} has to be minimized. These parameters are: two values of the orbital local moments $\mu_{\alpha}$ ($\alpha=1,2$), and two occupancies $n_{\alpha}$ of the orbitals. The total occupancy $n$ of the impurity can be adjusted by changing  the chemical potential. Here we fix $n=2$ (half-filling in the two-orbital case).

 Once the degeneracy is removed ($\delta \epsilon > J$) in the half-filled case both impurity electrons occupy the lower level and the total spin of the impurity is zero. The resulting local spectral densities show no quasi-particle peak at the Fermi energy (cf. Fig.~\ref{2_as_as_siam}). Fluctuations of the spin and orbital degrees of freedom are suppressed and the Kondo effect is absent.

In conclusion, the VLMA correctly provides approximate solutions to one- and two-orbital SIAM. Our results are in agreement with others \cite{hewson,cox_zawadowski} which shows that the VLMA is a good method for solving analogous magnetic impurity problems.

\section{Multi-orbital Hubbard model}

In this section we apply VLMA to solve the DMFT equations\cite{metzner89,dmft_review,dmft_phys_today} for multi-orbital systems. The Hamiltonian of the multi-orbital Hubbard model is given by\cite{hubbard64}
\begin{eqnarray}
H_{\rm{Hubbard}}=H_{kin} + H_U =  \nonumber \\
 H_{kin} + \frac{1}{2}\sum_{i,\alpha,\sigma} U_{\alpha}
n_{i\alpha\,{\sigma}} n_{i\alpha\,-\sigma} +\nonumber \\
 \frac{1}{2}\sum_{i,\sigma,\sigma'}\sum_{\alpha\neq\beta}\left( U'_{\alpha \beta}
-J_{\alpha \beta}\delta_{\sigma\sigma'}\right)
n_{i\alpha\sigma}n_{i\beta\sigma'},
\end{eqnarray}where the interaction part $H_U$ is local and identical to the interaction part of the Anderson impurity model. In the DMFT, the lattice model (in this case the Hubbard model) is mapped onto a multi-orbital impurity model with the same local interaction. The impurity is coupled to a bath of particles, whose properties are determined self-consistently. In order to solve the resulting equations, the single impurity problem has to be solved iteratively, in each step with a new hybridization function until self-consistency is reached. The VLMA can be used for arbitrary hybridization function and therefore it can serve as an impurity solver for the DMFT.

\subsection{Results for one and two-orbital Hubbard model}

\begin{figure}
\includegraphics[width=0.49\textwidth]{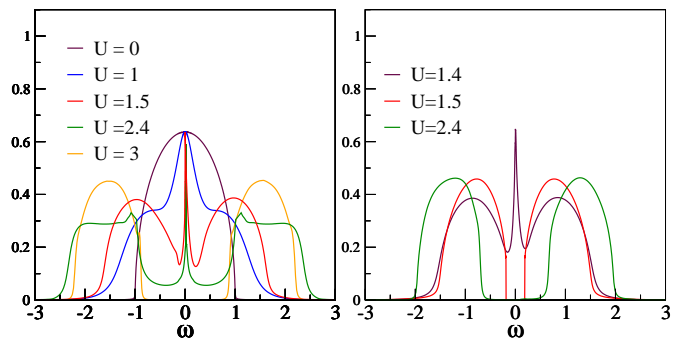}
\caption{(Color online). Local spectral functions for the one-orbital Hubbard model at half-filling. Left figure: the redistribution of the spectral weight with increasing $U$ due to Mott-Hubbard metal-insulator transition. Right figure illustrates hysteresis -- the solutions are obtained from initially insulating density of states} 
\label{art_mit2}
\end{figure}

The results of the DMFT+VLMA approach for the one-orbital Hubbard model on the Bethe lattice\cite{bethe_lattice1} with half-bandwidth $D=1$ are shown in Fig.~\ref{art_mit2}. The parameters here are: interaction $U_{\alpha=1}=U$, filling $n=1$, Fermi energy at $\tilde{\mu}=0$ and zero temperature. The VLMA solver for DMFT predicts a  Mott-Hubbard metal-insulator transition\cite{mott_mit} with  $U_{c1}\neq U_{c2}$ (hysteresis). In the current implementation of the method the value of $U_{c1}$ is obtained rather precisely ($U_{c1}=1.45$). In the case of $U_{c2}$, due to the exponential narrowing of the quasi-particle peak, the minimum of the ground state energy is increasingly difficult to find. That makes $U_{c2}$ harder to determine. The  obtained value $U_{c2} \gtrsim 2.5$ may be underestimated due to the difficulty in finding the minimum $\mu_V$ close to a divergence of the polarization propagator (as discussed in Sec.~IV).

We note that in the solution of the DMFT equations within the LMA the hysteresis was almost absent.\cite{eastwood_phd} This is due to the imposed Fermi liquid condition inside the LMA, which favors the metallic state. The VLMA is not biased in this respect. 

\begin{figure}
\includegraphics[width=0.49\textwidth]{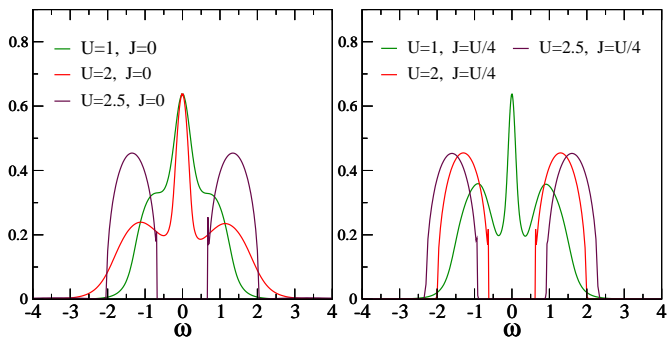}
\caption{(Color online). Local spectral functions for the two-orbital Hubbard model with degenerate orbitals. Left: the redistribution of the spectral weight with increasing $U$ due to Mott-Hubbard metal-insulator transition with no Hund's rule coupling $J=0$. Right: the metal-insulator transition in the case of $J\neq0$} 
\label{art_mit_a}
\end{figure}

The  Mott-Hubbard  metal-insulator transition in two-orbital case also shows hysteresis. In this case the transition is strongly influenced by the presence of non-zero exchange interaction $J_{\alpha\beta}=J$. The exchange interaction suppresses the Kondo effect and causes the transition to take place at a smaller value of the interaction $U_{\alpha}=U$ as shown in Fig.~\ref{art_mit_a} (here: $U'_{\alpha \beta} =U-2J$). 

In the two-orbital case, the DMFT solutions of the Hubbard model within VLMA show a metal-insulator transition that is orbital selective.\cite{kauch,kauch2} Such a transition is present both in the different bandwidths\cite{kauch2} case and in the non-degenerate orbitals case.\cite{kauch}

\section{Summary}

We have formulated in Sec.~III the variational local moment approach and have applied it in Sec.~IV to solve different versions of the single impurity Anderson problem. The VLMA is a comprehensive, thermodynamically consistent and conserving approximation, as we have shown in Sec.~IIIE. It can be used  for any values of the input parameters. Such a comprehensive method for solving the SIAM is a good method of choice for an efficient impurity solver in DMFT. We have presented in Sec.~V some results for the one- and two-orbital Hubbard model obtained within the DMFT with VLMA. 

The main point where the VLMA is different from the LMA of Logan {\it et al}.\cite{logan1,logan2,logan3,logan4} is that the lengths of local moments and orbital occupancies are determined variationally. Our method treats on equal footing both Fermi and non-Fermi liquid ground states as well as insulating, local moment states. 

The subtlety, however, in finding the ground state energy minimum at large interaction values, in particular close to the Mott-Hubbard metal-insulator transition, implies that at the moment the VLMA cannot yet be used as a "black-box" impurity solver. The semi-analytical formulation and implementation on the real-frequency axis make the method, in our opinion, valuable and we hope that the numerical difficulties of the current implementation will be overcome in the nearest future.

The method has been formulated for arbitrary temperature. However, it has not yet been implemented numerically in this regime. The potential difficulty at finite temperatures lies in an efficient calculation of the free energy, as mentioned in Sec.~IIID. The finite temperature numerical implementation may, however, be free from the numerical problem of finding the minimum close to a divergence, which we faced in the ground state case. These issues are also left for the nearest future.

\begin{acknowledgments}

We would like to thank  Dieter Vollhardt for numerous and invaluable discussions. This work was partly supported by the Sonderforschungsbereich 484 of the Deutsche Forschungsgemeinschaft (DFG). We also acknowledge support from the Polish Ministry of Science and Education grant N202026 32/0705.

\end{acknowledgments}

\appendix*

\section{Real frequency implementation of the RPE used in Sec.~IIIC}



We introduce the spectral representation $D^{A/B\,\sigma}_{\alpha}(\omega)$ for the UHF Green's functions \eqref{uhf_green} in the following way
{\small
$$
\tilde{G}^{A/B\,\sigma}_{\alpha}(i\omega_n)= \int_{-\infty}^{+\infty}\!\!\! d\omega' \frac{D^{A/B\,\sigma}_{\alpha}(\omega')}{i\omega_n -\omega'},
$$}where $D^{A/B\,\sigma}_{\alpha}(\omega)$ is thus given by 
{\small $$
D^{A/B\,\sigma}_{\alpha}(\omega)=-\frac{1}{\pi} \Im \tilde{G}^{A/B\,\sigma}_{\alpha}(\omega+i0^+).
$$}Then  the polarization propagator  $^0\Pi^{AA \,-\sigma\sigma}_{\alpha \beta}(i\nu_n)$ from \eqref{pi_zero} is expressed by the spectral functions $D^{A\,\sigma}_{\alpha}(\omega)$ as follows
{\small
\begin{equation*}
^0\Pi^{AA\,-\sigma\sigma}_{\alpha \beta}(i\nu_n)\!=\! -\! \frac{1}{\beta} \sum_{i\omega_m} \int \!\!\!\int_{-\infty}^{+\infty}\!\!\! \frac {d\omega'd\omega'' D^{A \,\sigma}_{\beta}(\omega')D^{A\,-\sigma}_{\alpha}(\omega'')}{(i\nu_n+i\omega_m-\omega')(i\omega_m-\omega'')}.
\end{equation*} }Performing the sum over Matsubara frequencies\cite{fetter} we obtain
{\small
\begin{eqnarray}
^0\Pi^{AA\,-\sigma\sigma}_{\alpha \beta}(i\nu_n)= \int \!\!\!\int_{-\infty}^{+\infty}\!\!\!d\omega'd\omega'' \frac{D^{A\,\sigma}_{\beta}(\omega') D^{A\,-\sigma}_{\alpha}(\omega'-\omega'')}{i\nu_n-\omega''} \times \nonumber \\
(f(\omega')-f(\omega'-\omega'')),\nonumber
\end{eqnarray}
}with  $f(\omega)\equiv(1+\exp(\beta \omega))^{-1}$ being the Fermi-Dirac distribution function.
If we introduce the spectral representation $^0\chi ^{AA\,\sigma\sigma'}_{\alpha\beta}(\omega)$ also for the polarization propagator 
{\small $$
^0\Pi^{AA\,\sigma\sigma'}_{\alpha \beta}(i\nu_n)= \int_{-\infty}^{+\infty}\!\!\! d\omega' \frac{^0\chi ^{AA\,\sigma\sigma'}_{\alpha \beta}(\omega')}{i\nu_n -\omega'},
$$}with
{\small $$
^0\chi ^{AA\,\sigma\sigma'}_{\alpha\beta}(\omega)=-\frac{1}{\pi} \Im {^0\Pi}^{AA\,\sigma\sigma'}_{\alpha\beta}(\omega + i0^+),
$$}we can express $^0\chi ^{AA\,-\sigma\sigma}_{\alpha\beta}(\omega)$ by integrals over real frequencies with the UHF spectral functions in the following way
{\small
\begin{equation*}
^0\chi ^{AA\,-\sigma\sigma}_{\alpha\beta}(\!\omega)\!=\!\int_{-\infty}^{+\infty} \!\!\!\! d\omega' D^{A\,\sigma}_{\beta}(\omega') D^{A\,-\sigma}_{\alpha}(\!\omega'
\!-\!\omega)(f(\omega')\!-\!f(\omega'-\omega)).
\end{equation*}}We can analogously express the other (orbital)  polarization propagator $^0\Pi^{AA \,\sigma\sigma}_{\alpha \beta}(i\nu_n)$ by integrals over the UHF spectral functions. The real parts of the polarization propagators can be found by using Kramers-Kronig relations.\cite{fetter}

The sum over the Matsubara frequencies in Eq.~\eqref{selfenergia3} is done with the use of the spectral representations of the full polarization propagator  $\mathbf{\Pi}^{AA}(i\nu_n)$, i.e.
{\small
$$
\Pi^{AA\,\sigma\sigma'}_{\alpha \beta}(i\nu_n)= \int_{-\infty}^{+\infty}\!\!\! d\omega' \frac{\chi ^{AA\,\sigma\sigma'}_{\alpha \beta}(\omega')}{i\nu_n -\omega'},
$$}where
{\small $$
\chi ^{AA\,\sigma\sigma'}_{\alpha\beta}(\omega)=-\frac{1}{\pi} \Im {\Pi}^{AA\,\sigma\sigma'}_{\alpha\beta}(\omega + i0^+).
$$}Then Eq.~\eqref{selfenergia3} can be rewritten to give
{\small
$$
\Sigma^{A}_{\alpha \, \sigma}(i\omega_n)\!=\!\frac{1}{\beta}\! \!\!\sum_{\beta \, \sigma'\,i\nu_m}\!\! (U_{\alpha \beta}^{\sigma \sigma'})^2 \!\!\int \!\!\!\int_{-\infty}^{+\infty}\!\!\frac {d\omega'd\omega''D^{A\,\sigma'}_{\beta}(\omega')\chi^{AA\,\sigma'\sigma}_{\beta \alpha}(\omega'')}{(i\omega_n-i\nu_m-\omega')(i\nu_m-\omega'')}. 
$$}After performing the sum over Matsubara frequencies we arrive at the expression
{\small
\begin{eqnarray}
\Sigma^{A}_{\alpha \, \sigma}(i\omega_n)=\sum_{\beta \, \sigma'} (U_{\alpha \beta}^{\sigma \sigma'})^2 \,  \int \!\!\!\int_{-\infty}^{+\infty}\!\!\!d\omega'd\omega'' \frac{D^{A\,\sigma'}_{ \beta}(\omega''-\omega')}{i\omega_n-\omega''} \times \nonumber \\ 
 \chi^{AA\,\sigma'\sigma}_{\beta \alpha}(\omega') (\tilde{f}(\omega'-\omega'')-f(\omega')), \nonumber
\end{eqnarray} }which contains the Fermi-Dirac distribution  $f(\omega)$ together with a function $\tilde{f}(\omega)\equiv(1-e^{\beta \omega})^{-1}$. Introducing the spectral representation also for the self-energy
{\small $$
\Sigma_{\alpha}^{A\sigma}(i\omega_n)=\int_{-\infty}^{+\infty} \!\!\! d\omega\frac{B_{\alpha}^{A\sigma}(\omega)}{i\omega_n-\omega}
$$}with
{\small$$
B_{\alpha}^{A\sigma}(\omega)=-\frac{1}{\pi} \Im \Sigma_{A\alpha}^{\sigma}(\omega + i0^+),
$$}we are able to calculate the self-energy using only integrals over real frequencies
{\small
\begin{eqnarray}
B_{\alpha}^{A\,\sigma}(\omega)=  \sum_{\beta \, \sigma'} (U_{\alpha \beta}^{\sigma \sigma'})^2  \int_{-\infty}^{+\infty}\!\!\!d\omega' D^{A\sigma'}_{\beta}(\omega-\omega')  \times \nonumber \\ 
 \chi^{AA \,\sigma'\sigma}_{\beta \alpha}(\omega')(\tilde{f}(\omega'-\omega)-f(\omega')). \nonumber
\label{eq:sigma}
\end{eqnarray}
}The real part of self-energy can be obtained from the imaginary part with the use of Kramers-Kronig relations.

\end{document}